# Algorithm for generating irreducible site-occupancy configurations

Ji-Chun Lian[1], Hong-Yu Wu[1], Wei-Qing Huang[1*], Wangyu Hu[2] and Gui-Fang Huang[1#]

[1]Department of Applied Physics, School of Physics and Electronics, Hunan University, Changsha 410082, China

[2]School of Materials Science and Engineering, Hunan University, Changsha 410082, China

Generating irreducible site-occupancy configurations by taking advantage of crystal symmetry is a ubiquitous method for accelerating of disordered structure prediction, which plays an important role in condensed matter physics and material science. Here, we present a new algorithm for generating irreducible site-occupancy configurations, that works for arbitrary parent cell with any supercell expansion matrix, and for any number of atom types with arbitrary stoichiometry. The new algorithm identifies the symmetrically equivalent configurations by searching the space group operations of underlying lattice and building the equivalent atomic matrix based on it. Importantly, an integer representation of configurations can greatly accelerate the speed of elimination of duplicate configurations, resulting into a linear scale of run time with the number of irreducible configurations that finally found. Moreover, based on our new algorithm, we write the corresponding code named as *disorder* in FORTRAN programming language, and the performance test results show that the time efficiency of our *disorder* code is superior to that of other related codes (*supercell*, *enumlib* and SOD).

---

*. Corresponding author. *E-mail address*: wqhuang@hnu.edu.cn

#. Corresponding author. *E-mail address*: gfhuang@hnu.edu.cn

# I. INTRODUCTION

Searching the unknown structures is of paramount potential impact in condensed matter physics and materials science [1-4]. The high efficiency and accuracy of first-principles calculations in describing the structural thermodynamic stability [5-7] and other related properties [8-10] has greatly promoted the process of structure prediction — extensive structural ground state properties calculations become reality. However, the evaluating of structural relative stabilities based on first-principles calculations becomes challenging for the issue of site-occupancy disorder [11-15] — a subset of structure prediction problem, whose structures are derived from a specific underlying lattice. The reason is that the underlying lattice is often a large supercell, which not only increases the computational load of first-principles, but also the number of structures derived from it dramatically increases due to combinatorial explosion. Nevertheless, the efforts have not been stopped and will never be stopped, as many profound physical and material phenomena are observed in disordered structures, and bring about important applications in fields such as metallic alloys [16,17], non-stoichiometric materials [18,19] and high-temperature superconductors [20-23]. Aiming to overcome this obstacle, several methods have been developed to reduce the generated atomic configurations [24-27]. In this paper, we focus on one of those methods, i.e., searching the irreducible configurations in a complete list of atomic configurations by taking advantage of crystal symmetry.

Many combinatorially distinct configurations are geometrically identical — they are related by the symmetry operations of underlying lattice. The main idea of searching irreducible configurations is to generate an exhaustive list of combinatorially distinct configurations, then eliminating the duplicates, i.e. the symmetrically equivalent configurations. Such an idea is adopted by many existing algorithms [24,28-30], but the implementations are diverse from each other. In the following, several representative algorithms for generating irreducible site-occupancy configurations will be briefly introduced.

The algorithm implemented in SOD software (released in 2007) is one of the excellent representatives [24]. The SOD algorithm identifies the equivalent configurations utilizing space group operations of the underlying lattice, which are reading from a database of space group operations. The superiority of SOD algorithm lies in its simplicity in concept and programming. However, it only works for the binary site-occupancy systems, and the non-diagonal supercell expansion matrix is forbidden in SOD software because of programming defects. What's worse, its run time will grow explosively with the increase of the number of configurations.

Compared with SOD algorithm, an extraordinary progress is made by the algorithm implemented in *enumlib* software (released in 2008) [28]. The *enumlib* algorithm applies to any parent cell, arbitrary supercell expansion matrix and multinary (binary, ternary, quaternary etc.) site-occupancy systems. For the first version, the *enumlib* algorithm applies only to the Bravais lattices, and it not works at a certain stoichiometry. These problems have been solved in the later two extensions [31,32]. The key concept of the *enumlib* algorithm is to use the quotient group associated with the underlying lattice and an integer representation of the configurations to determine all unique structures. Profit by this concept, the *enumlib* algorithm is orders of magnitude faster than SOD algorithm and realized a linear scale of run time with the number of irreducible configurations, which is the best possible scaling for this type of problem.

Although the run time of the *enumlib* algorithm scales linearly with the number of irreducible configurations, the calculation time is far beyond a reasonable limit when the number of configurations is huge (say millions). Therefore, the algorithm performance still has the possibility to be further improved, which has become a reality in *supercell* software (released in 2016) [30]. The algorithm implemented in *supercell* is similar in essence to the *enumlib* algorithm, but its time efficiency is about ten times that of *enumlib* algorithm. However, the *supercell* algorithm can not support non-diagonal supercell expansion matrix, which limits its exploration for different supercell shapes.

In this paper, we present a new implementation for the generation algorithm of irreducible site-occupancy configurations. In our algorithm, we enumerate the combination configurations to represent the combinatorially distinct atomic configurations, and extend it to the multinary site-occupancy systems. Our new algorithm identifies the symmetrically equivalent configurations by an equivalent atomic matrix, while the space group operations used to build it are searched from the structural information (only the lattice parameters and atomic positions are needed) of arbitrary supercell straightforwardly. As a result, the new algorithm is applicable for arbitrary parent cell with any supercell expansion matrix including the non-diagonal one, which is forbidden in SOD and *supercell* softwares. In addition, a linear scale of run time with the number of irreducible configurations is achieved also, benefit by the concept of using a series of consecutive integers to represent the configurations. Most of all, the speed of eliminating duplicate configurations is greatly accelerated, due to the efficient conversion algorithm of configurations to integers, and hence faster than that of *supercell* algorithm.

**II. METHODOLOGY**

The algorithm for generating irreducible site-occupancy configurations can be summarized into four critical tasks: (1) searching space group operations to build the equivalent atoms matrix, (2) enumerating all atomic configurations based on combinatorics, (3) eliminating the duplicate configurations by using symmetries (the most time-consuming procedure), (4) converting atomic configurations into consecutive integers to accelerate the third task. In the following, a detailed description of the algorithm implementation is presented.

**A. Searching space group operations**

Searching space group operations is a key point for identifying equivalent atomic configurations. The algorithm for searching space group operations is based on *Spglib*: a software library for crystal symmetry search [33]. Although the original algorithm implemented in *Spglib* is used to the primitive cell, it can be applied to arbitrary cell with appropriate modifications. The algorithm of searching space group operations used here is a simplified and modified version of *Spglib*.

The space group operation is defined by a 3×3 integer matrix $W$ (rotation part) and a 3×1 decimal column vector $\omega$ (translation part). An arbitrary atomic point $x$ on the fractional coordinates must be sent to another atomic point (or itself) $x'$ with the same atomic type by one of the valid space group operations $(W, \omega)$ by $x' = Wx + \omega$.

In the following, a brief outline of the algorithm for searching all valid space group operations is presented (the elaborated description is given in Ref. [33]).

***1. Searching pure translation operations*** (See Ref. [33], Step (a))

The pure translation operations are expressed as $(I, \omega)$, where $I$ is the identity matrix. Generally, the input cell (i.e., the underlying lattice) is a non-primitive cell, so that there are multiple pure translation operations $(I, \omega)$ will be found, and the number ($N_t$) of pure translation operations is the volume of the input cell divided by the volume of its primitive cell.

***2. Searching lattice point group operations*** (See Ref. [33], Step (f))

In this step, all the matrix elements of possible $W$ are selected from {-1, 0, 1} to satisfy $|\det(W)| = 1$. Afterwards, possible $W$ are further screened by the conditions for metric tensor $G$, which is defined by $G = (a, b, c)^T (a, b, c)$, where $(a, b, c)$ is the lattice basis vectors. In the *spglib*, the basis vectors used to define metric tensor $G$ is that of primitive cell. Here, we replace it with that of the input cell to obtain the space group operations of arbitrary cell.

***3. Searching space group operations*** (See Ref. [33], Step (g))

For a non-primitive cell, similar to pure translation operations, one $W$ will correspond to multiple $\omega$. Thus, the total number ($N_s$) of space group operations is the number ($N_r$) of its rotation part multiply by the number ($N_t$) of its translation part, i.e., $N_s = N_r \times N_t$. In reality, for one $W$ only one (any one) corresponding $\omega$ need to be searched out, while the full space group operations can be obtained by combining pure translation operations.

*4. Building equivalent atomic matrix*

Building equivalent atomic matrix will advantageous to identify the equivalent atomic configurations conveniently and quickly. A set of atomic points labeled with consecutive integers are transformed into another set of atomic points by a space group operation. The labels of the transformed atomic points constitute one row elements of the equivalent atomic matrix, while the complete equivalent atomic matrix can be obtained by traversing all space group operations.

**B. Enumerating all atomic configurations**

The enumeration of all possible atomic configurations is based on combinatorics. Consider a underlying lattice with one disordered crystallographic site and $K$ atoms, let $N$ different types of atoms (the vacancies are deemed as a special type of atoms) occupying this site and the atomic number of each type is $k_i$ ($i = 1 \ldots N$), satisfying $K = \sum_{i=1}^{N} k_i$. The total number ($N_c$) of atomic configurations can be obtained by a generalized combination number formula:

$$N_c(k_1, k_2, \ldots, k_N) = \frac{(k_1 + k_2 + \ldots + k_N)!}{k_1! k_2! \ldots k_N!} = \frac{\left(\sum_{i=1}^{N} k_i\right)!}{\prod_{i=1}^{N} k_i!}. \tag{1}$$

For clarity, we will introduce our new algorithm in two steps. Specifically, we start with a special case, i.e. the binary ($N = 2$) site-occupancy case, and then, extend it to the general case, i.e. the multinary ($N \geq 2$) site-occupancy case.

*1. Binary site-occupancy*

The binary site-occupancy is two different types of atoms occupying one disordered crystallographic site. In such a special case, according to **Eq. (1)**, the total number of site-occupancy configurations is expressed as:

$$N_c(k_1,k_2) = \frac{(k_1+k_2)!}{k_1!k_2!} = \frac{K!}{k_1!k_2!} = C_K^{k_1} = C_K^{k_2}, \tag{2}$$

i.e. the common combination number formula, where $C_n^m = \frac{n!}{m!(n-m)!}$.

Hereafter, for convenience, we use different colors to represent different types of atoms. As mentioned above, our task is enumerating all possible atomic configurations. The algorithm enumerates all combination configurations to represent the atomic configurations, which is expressed as a list **A**, with $C_K^{k_1}$ rows and $k_1$ columns. Obviously, we only need to determine the atomic configurations of one color, while the others occupy the remaining atomic positions. Therefore, there are two ways to enumerate site-occupancy configurations, as shown in Fig. 1, where an example is for $K = 6$, $k_1 = 4$ (red), $k_2 = 2$ (blue). In practice, undoubtedly, we can enumerate the lesser one of the two colors to reduce the computational load.

## 2. Extend to multinary site-occupancy

The multinary site-occupancy is $N$ ($N \geq 2$) different types of atoms occupying one disordered crystallographic site. The total number of possible atomic configurations for the multinary site-occupancy can be obtained from **Eq. (1).** Here we rewrite it as:

$$N_c(k_1,k_2,...,k_N) = \frac{K!}{k_1!k_2!...k_N!} = C_K^{k_1} C_{K-k_1}^{k_2} C_{K-k_1-k_2}^{k_3} ... C_{K-k_1-k_2-...-k_{N-2}}^{k_{N-1}}. \tag{3}$$

Its physical meaning is obvious, that is let $k_1$ atoms occupying $K$ atomic positions, then let $k_2$ atoms occupying the remaining $K - k_1$ atomic positions, and so on. Of course, the order of $k_1, k_2, ..., k_N$ is irrelevant. This means that we can decompose the multinary site-occupancy into several binary site-occupancy.

We represent the $n$-th binary combination configurations in **Eq. (3)** as a list **B**$^{(n)}$, with $C_{K-k_1-k_2-...-k_{n-1}}^{k_n}$ rows and $k_n$ columns. Then, the complete combination configurations **B** can be defined by using these binary combination configurations, as follows:

$$B_n = B_i^{(1)} \bigcup B_j^{(2)} \bigcup ... \bigcup B_k^{(N-1)}$$

$$n \in \left[1, N_c(k_1, k_2,..., k_N)\right],\ i \in \left[1, C_K^{k_1}\right],\ j \in \left[1, C_{K-k_1}^{k_2}\right],\ k \in \left[1, C_{K-k_1-k_2-...-k_{N-2}}^{k_{N-1}}\right]$$

However, unlike the binary site-occupancy, the combination configurations defined above are not the atomic configurations. Therefore, we need to convert the combination configurations to the atomic configurations.

Given configuration **a**, and its corresponding combination configuration **b** can be disassembled into follows:

$$a = a^{(1)} \bigcup a^{(2)} \bigcup ... \bigcup a^{(N-1)}$$
$$b = b^{(1)} \bigcup b^{(2)} \bigcup ... \bigcup b^{(N-1)}$$

Moreover, a series of arrays $\left(L^{(1)}, L^{(2)},..., L^{(N-1)}\right)$ with $\left(K, K-k_1,..., K-k_1-...-k_{N-2}\right)$ columns are used to labeling the atomic positions, where $L^{(1)}$ is the original atomic labels, $L^{(2)}$ is the atomic labels after removing $a^{(1)}$, and the like, $L^{(N-1)}$ is the atomic labels after removing $a^{(1)} \bigcup a^{(2)} \bigcup ... \bigcup a^{(N-2)}$. Then, the transformation relationship between **a** and **b** can be written as:

$$a_i^{(n)} = L_{b_i^{(n)}}^{(n)}, \tag{4}$$

where, $n \in [1, N-1]$, $i \in [1, k_n]$, and $i$ is the column index of $a^{(n)}$ or $b^{(n)}$. Here, we take the ternary site-occupancy as an example, and specify $K = 6$, $k_1 = 2$ (red), $k_2 = 3$ (blue), $k_3 = 1$ (green). For the ternary site-occupancy, we can decompose it into two binary site-occupancy, the combination configurations of them and the resulting atomic configurations are shown in **Fig. 2**.

**C. Converting configurations to integers**

As stated previously, the third task is eliminating the duplicate configurations, which is the most time-consuming procedure. The process of eliminating duplicate configurations can be greatly accelerated by the directly determination of configuration indexes, which is expressed as a set of consecutive integers. In the following, we will show how it is implemented and explain why it is so efficient.

*1. Binary site-occupancy*

For the binary site-occupancy, the configuration index of arbitrarily atomic configuration can be calculated by the following procedure. Firstly, we build a matrix **D** as follows:

$$D_{ij} = \begin{cases} 0 &, i = 1 \\ C_{k_1-j+i-2}^{k_1-j} &, i \neq 1 \end{cases},$$

where, the number of rows is $K - k_1 + 1$, the number of columns is $k_1$, $i$ ($j$) is the row (column) index, and we define $C_n^0 = 1$. Based on matrix **D**, another matrix **E** with the same size can be obtained, as shown below:

$$E_{ij} = \sum_{k=1}^{i} D_{kj}. \tag{5}$$

Then, given a configuration **a**, its configuration index $m$ can be expressed as:

$$m = C_K^{k_1} - \sum_{j=1}^{k_1} E_{ij}, \tag{6}$$

where $i = A_{-1\,j} - a_j + 1$, $A_{-1}$ is the last configuration in list **A**. Again, we specify $K = 6$, $k_1 = 4$ (red), $k_2 = 2$ (blue) as an example, and the calculation flow is shown in **Fig. 3**.

*2. Extend to multinary site-occupancy*

For the multinary site-occupancy, similar to matrix **E** (see **Eq. (5)**) of the binary site-occupancy, a set of matrices $\mathbf{E}^{(n)}$ corresponding to the $n$-th combination configurations can be built using the same method. Actually, however, these $N-1$ matrices are sub-matrices of a larger matrix, meaning that we only need to construct one matrix **E** whose number of rows and columns are the maximum of that in matrices $\mathbf{E}^{(n)}$. Afterwards, given configuration **a**, its configuration index $m$ can be expressed as:

$$m = N_c - \frac{N_c}{C_K^{k_1}} \sum_{j_1=1}^{k_1} E_{i_1 j_1}^{(1)} - \frac{N_c}{C_K^{k_1} C_{K-k_1}^{k_2}} \sum_{j_2=1}^{k_2} E_{i_2 j_2}^{(2)} - \ldots - \sum_{j_{N-1}=1}^{k_{N-1}} E_{i_{N-1} j_{N-1}}^{(N-1)},$$

where, $N_c = \dfrac{K!}{k_1! k_2! \ldots k_N!}$, $i_n = B^{(n)}_{-1 j_n} - b^{(n)}_{j_n} + 1$, $n \in [1, N-1]$. Therefore, we need to convert the atomic configuration **a** into the corresponding combination configuration **b**, and it is not difficult to realize by the inverse process of **Eq. (4)**. In addition, one can notice that when $N = 2$, the above formula will degenerate into the binary site-occupancy case (see **Eq. (6)**). Finally, we specify $K = 6$, $k_1 = 2$ (red), $k_2 = 3$ (blue), $k_3 = 1$ (green) as an example, and the calculation flow is shown in **Fig. 4**.

### D. Eliminating duplicate configurations

Our new algorithm used to eliminate duplicate configurations is illustrated by a flowchart as shown in **Fig. 5**. The preprocessing algorithm (the red frame) involves the obtaining of space group operations, equivalent atomic matrix, all binary combination configurations, and matrix **E**. Afterwards, a parent-loop (the blue frame) traversing all atomic configurations, and a sub-loop (the green frame) traversing all space group operations are performed. In the parent-loop, if the atomic configuration has been eliminated, then skip it directly, else add it to the list **I** (represents the irreducible configurations), and entries the sub-loop. In the sub-loop, the space group operation, including rotation, translation, and permutation, maps a configuration **a** to another configuration (or itself) **a'**. If the configuration **a'** has been eliminated, then skip it again, else eliminate it. Obviously, the key point of the above process is how to eliminate the atomic configuration, and how to know a configuration has been eliminated. This is achieved by a logical list **U** corresponding to the atomic configurations one-by-one. At the beginning, all elements in **U** are setting to "**TRUE**", which means that all configurations have not been eliminated. If a configuration needs to be eliminated, set the corresponding element in **U** to "**FALSE**". All in all, the fast conversion of configurations to integers plays a pivotal role for the improvement of algorithm performance.

## III. NEW ALGORITHM PERFORMANCE AND SUPERIORITY

Based on aforementioned algorithm, we write the corresponding code named as *disorder* in FORTRAN programming language. After that, an example is given to test the algorithm performance, and compared with those algorithms implemented in *supercell*, *enumlib* and SOD softwares.

The example is based on a face-centered cubic (FCC) parent lattice, as many structures of inter-metallic compounds are derived from it. The point group of FCC lattice is $O_h$, with 48 point group operations (rotation parts of space group operations), which is the highest symmetry. Moreover, the volume of FCC unit cell (4 atoms) is 4 times of its primitive cell (1 atoms), which means that the FCC unit cell possesses 4 pure translation operations. For a FCC unit cell, therefore, the total number of space group operations is 48×4 = 192 (48 rotations and 4 translations). In our example, a 2×2×2 supercell (32 atoms) is adopted to test the algorithm performance. This is a challenging testing, because its 48×4×8 = 1536 space group operations (48 rotations and 4×8 translations) is really high.

We enumerate all combinatorially distinct stoichiometries for binary site-occupancy and present the calculation results (containing total configurations, irreducible configurations and run time) in **Table I**. One can see that all four codes, including *disorder* (this work), *supercell*, *enumlib* and SOD, give the same number of total and irreducible configurations, which proves the correctness of our algorithm. However, the run time of the four codes is different: the run time of SOD code grows sharply with the increase of the number of irreducible configurations, but that of other three codes grow slowly.

For the convenience of comparing the run time of *disorder*, *supercell* and *enumlib* codes, we plot the run time as a function of the number of irreducible configurations, as shown in **Fig. 6**. One can notice that a linear relationship between run time and the number of irreducible configurations is appeared when the number of irreducible configurations is greater than a critical value, for all the three codes. Such a critical value depends on the ratio of preprocessing time to total time: a linear scale is emerged when the ratio is small enough, i.e. the preprocessing time is negligible compared with the total time. The large critical value of *supercell* is owing to its long preprocessing time primarily, while the small critical value of *enumlib* is mainly due to its long total time. For our *disorder* code, both the total time and preprocessing time are relatively small, so that its critical value is between that of *enumlib* and *supercell*. Moreover, we can see that our *disorder* code has the lowest run time throughout the stoichiometric range (see **Fig. 6**), and the cumulative run time (see last row of **Table I**) of *disorder* is 18 minutes, about half that of *supercell* (35 minutes) and less than one twenty-third that of *enumlib* (7 hours), which indicates the high efficiency of our new algorithm. As a summary, a brief comparisons of *disorder*, *supercell*, *enumlib* and SOD codes are presented in **Table II**.

## IV. SUMMARY

We have developed a new algorithm for generating irreducible site-occupancy configurations. The new algorithm processes the multinary site-occupancy problem by decomposing it into several binary site-occupancy. Afterwards, based on combinatorics, the algorithm enumerates all binary combination configurations, and converts them to atomic configurations. In the procedure of eliminating duplicate configurations, the algorithm identifies the duplicate configurations by an equivalent atomic matrix, which is built from the space group operations of underlying lattice. In our algorithm, the space group operations are searched from the structural information of underlying lattice directly, and works for arbitrary parent cell with any supercell expansion matrix. Most important, an efficient conversion of configurations to integers is adopted to accelerate the process of eliminating duplicate configurations, and finally realized a linear scale of run time with the number of irreducible configurations, which is proved by the performance testing in a 2×2×2 FCC lattice. Moreover, the results also indicate that the time efficiency of the new algorithm is greatly superior to other algorithms with the same or different time complexity.

## ACKNOWLEDGMENTS

We acknowledge financial support from National Natural Science Foundation of China (Grant Nos. 51772085 and U1830138)

# References

[1] R. Caputo and A. Tekin, Ab-initio crystal structure prediction. A case study: NaBH4, J. Solid State Chem. **184**, 622 (2011).

[2] R. Podeszwa, B. M. Rice, and K. Szalewicz, Predicting structure of molecular crystals from first principles, Phys. Rev. Lett. **101**, 115503 (2008).

[3] Y. Wang, J. Lv, L. Zhu, and Y. Ma, Crystal structure prediction via particle-swarm optimization, Phys. Rev. B **82**, 094116 (2010).

[4] S. M. Woodley and R. Catlow, Crystal structure prediction from first principles, Nat. Mater. **7**, 937 (2008).

[5] K. Hu, Q.-J. Chen, and S.-Y. Xie, Pressure induced superconductive 10-fold coordinated TaS2: a first-principles study, J. Phys.-Condens. Mat. **32**, 085402 (2020).

[6] K. Hu, J. Lian, L. Zhu, Q. Chen, and S.-Y. Xie, Prediction of Fe2P-type TiTe2 under pressure, Phys. Rev. B **101**, 134109 (2020).

[7] A. J. Misquitta, G. W. A. Welch, A. J. Stone, and S. L. Price, A first principles prediction of the crystal structure of C6Br2ClFH2, Chem. Phys. Lett. **456**, 105 (2008).

[8] P. A. Korzhavyi, I. A. Abrikosov, B. Johansson, A. V. Ruban, and H. L. Skriver, First-principles calculations of the vacancy formation energy in transition and noble metals, Phys. Rev. B (Condensed Matter) **59**, 11693 (1999).

[9] J.-C. Lian, W.-Q. Huang, W. Hu, and G.-F. Huang, Electrostatic Potential Anomaly in 2D Janus Transition Metal Dichalcogenides, Ann. Phys. (Berlin) **531**, 1900369 (2019).

[10] X. Y. Zhao and D. Vanderbilt, First-principles study of structural, vibrational, and lattice dielectric properties of hafnium oxide, Phys. Rev. B **65**, 233106 (2002).

[11] P. Soven, Contribution to the theory of disordered alloys, Phys. Rev. **178**, 1136 (1969).

[12] J. Biasco, J. Garcia, J. M. de Teresa, M. R. Ibarra, J. Perez, P. A. Algarabel, C. Marquina, and C. Ritter, Structural, magnetic, and transport properties of the giant magnetoresistive perovskites La2/3Ca1/3Mn1-xAlxO3-delta, Phys. Rev. B: Condens. Matter **55**, 8905 (1997).

[13] A. Landa, P. Soderlind, A. Ruban, L. Vitos, and L. Pourovskii, First-principles phase diagram of the Ce-Th system, Phys. Rev. B **70**, 224210 (2004).

[14] R. Grau-Crespo, N. H. de Leeuw, and C. R. A. Catlow, Cation distribution and magnetic ordering in FeSbO4, J. Mater. Chem. **13**, 2848 (2003).

[15] V. Shuvaeva, Y. Azuma, K. Yagi, H. Terauchi, R. Vedrinski, V. Komarov, and H. Kasatani, Ti off-center displacements in Ba1-xSrxTiO3 studied by EXAFS, Phys. Rev. B **62**, 2969 (2000).

[16] J. Guevara, V. Vildosola, J. Milano, and A. M. Llois, Half-metallic character and electronic properties of inverse magnetoresistant Fe1-xCoxSi alloys, Phys. Rev. B **69**, 184422 (2004).

[17] K. Ozdogan, E. Sasioglu, B. Aktas, and I. Galanakis, Doping and disorder in the Co2MnAl and Co2MnGa half-metallic Heusler alloys (vol 74, 172412, 2006), Phys. Rev. B **80**, 029901 (2009).

[18] E. Koga, H. Moriwake, K.-I. Kakimoto, and H. Ohsato, Raman spectroscopic evaluation and microwave dielectric property of order/disorder and stoichiometric/non-stoichiometric Ba(Zn1/3Ta2/3)O-3, Ferroelectrics **356**, 438 (2007).

[19] A. A. Rempel, A. I. Gusev, and M. Y. Belyaev, 93Nb NMR study of an ordered and a disordered non-stoichiometric niobium carbide, J. Phys. C **20**, 5655 (1987).

[20] Chudnovsky, Hexatic vortex glass in disordered superconductors, Phys. Rev. B: Condens. matter **40**, 11355 (1989).

[21] S. A. Davydov *et al*., Effects of localisation in atomic-disordered high-Tc superconductors, Int. J. Mod. Phys. B **3**, 87 (1989).

[22] I. Grosu, Veres, and Crisan, Decrease in critical temperature due to disorder and magnetic correlations in two-dimensional superconductors, Phys. Rev. B: Condens. matter **50**, 9404 (1994).

[23] Williams, Kwei, R. B. Von Dreele, Raistrick, and Bish, Joint x-ray and neutron refinement of the structure of superconducting YBa2Cu, Phys. Rev. B: Condens. matter **37**, 7960 (1988).

[24] R. Grau-Crespo, S. Hamad, C. R. A. Catlow, and N. H. de Leeuw, Symmetry-adapted configurational modelling of fractional site occupancy in solids, J. Phys.-Condens. Mat. **19**, 256201 (2007).

[25] C. E. Mohn and W. Kob, A genetic algorithm for the atomistic design and global optimisation of substitutionally disordered materials, Comp. Mater. Sci. **45**, 111 (2009).

[26] J. A. Purton, M. Y. Lavrentiev, and N. L. Allan, Monte Carlo simulation of GaN/InN mixtures, J. Mater. Chem. **15**, 785 (2005).

[27] I. T. Todorov, N. L. Allan, M. Y. Lavrentiev, C. L. Freeman, C. E. Mohn, and J. A. Purton, Simulation of mineral solid solutions at zero and high pressure using lattice statics, lattice dynamics and Monte Carlo methods, J. Phys.-Condens. Mat. **16**, S2751 (2004).

[28] G. L. W. Hart and R. W. Forcade, Algorithm for generating derivative structures, Phys. Rev. B **77**, 224115 (2008).

[29] S. Mustapha, P. D'Arco, M. De La Pierre, Y. Noel, M. Ferrabone, and R. Dovesi, On the use of symmetry in configurational analysis for the simulation of disordered solids, J. Phys.-Condens. Mat. **25**, 105401 (2013).

[30] K. Okhotnikov, T. Charpentier, and S. Cadars, Supercell program: a combinatorial structure-generation approach for the local-level modeling of atomic substitutions and partial occupancies in crystals, J. Cheminformatics **8**, 17 (2016).

[31] G. L. W. Hart and R. W. Forcade, Generating derivative structures from multilattices: Algorithm and application to hcp alloys, Phys. Rev. B **80**, 014120 (2009).

[32] G. L. W. Hart, L. J. Nelson, and R. W. Forcade, Generating derivative structures at a fixed concentration, Comp. Mater. Sci. **59**, 101 (2012).

[33] A. Togo, I. Tanaka, *Spglib*: a software library for crystal symmetry search, arXiv:1808.01590 (2018).

**Tables**

**TABLE I**. The calculation results of *disorder* (this work), *supercell*, *emumlib* and SOD codes for all combinatorially distinct stoichiometries in a 2×2×2 FCC lattice.

| Stoichiometries | Total configurations | Irreducible configurations | Run time (sec) | | | |
|---|---|---|---|---|---|---|
| | | | *disorder* | *supercell* | *enumlib* | SOD |
| 1:31 | 32 | 1 | 0.0218 | 13.354 | 0.0745 | 0.0681 |
| 2:30 | 496 | 5 | 0.0207 | 13.542 | 0.1517 | 0.0899 |
| 3:29 | 4960 | 14 | 0.0226 | 13.746 | 0.3102 | 0.5776 |
| 4:28 | 35960 | 71 | 0.0385 | 13.565 | 1.0856 | 21.173 |
| 5:27 | 201376 | 223 | 0.0812 | 13.721 | 3.2949 | 451.29 |
| 6:26 | 906192 | 874 | 0.2459 | 14.304 | 12.447 | 10248 |
| 7:25 | 3365856 | 2706 | 0.7622 | 16.553 | 39.866 | > 2 days |
| 8:24 | 10518300 | 8043 | 2.2275 | 23.266 | 119.01 | / |
| 9:23 | 28048800 | 20123 | 6.9478 | 36.764 | 300.72 | / |
| 10:22 | 64512240 | 45497 | 18.398 | 65.990 | 701.43 | / |
| 11:21 | 129024480 | 88716 | 41.692 | 116.29 | 1377.3 | / |
| 12:20 | 225792840 | 154379 | 81.531 | 191.36 | 2333.6 | / |
| 13:19 | 347373600 | 234803 | 137.11 | 286.38 | 3559.0 | crashed |
| 14:18 | 471435600 | 318348 | 209.04 | 379.59 | 4857.5 | crashed |
| 15:17 | 565722720 | 379926 | 283.44 | 454.09 | 5836.6 | crashed |
| 16:16 | 601080390 | 404582 | 306.24 | 473.10 | 6123.4 | crashed |
| Cumulative | 2248023842 | 1658311 | 1087.7 | 2125.7 | 25264 | / |

**TABLE II**. The comparisons of *disorder* (this work), *supercell*, *emumlib* and SOD codes. The performance is the cumulative time of all combinatorially distinct stoichiometries in a 2×2×2 FCC lattice.

|  | *disorder* | *supercell* | *enumlib* | SOD |
|---|---|---|---|---|
| Public release | 2020 | 2016 | 2008 | 2007 |
| Programming language | FORTRAN | C++ | FORTRAN | FORTRAN |
| Non-diagonal supercell expansion matrix | yes | no | yes | no |
| Arbitrary number of atom types (i.e. the multinary systems) | yes | yes | yes | no |
| Performance | 18 min | 35 min | 7 h | / |

**Figures**

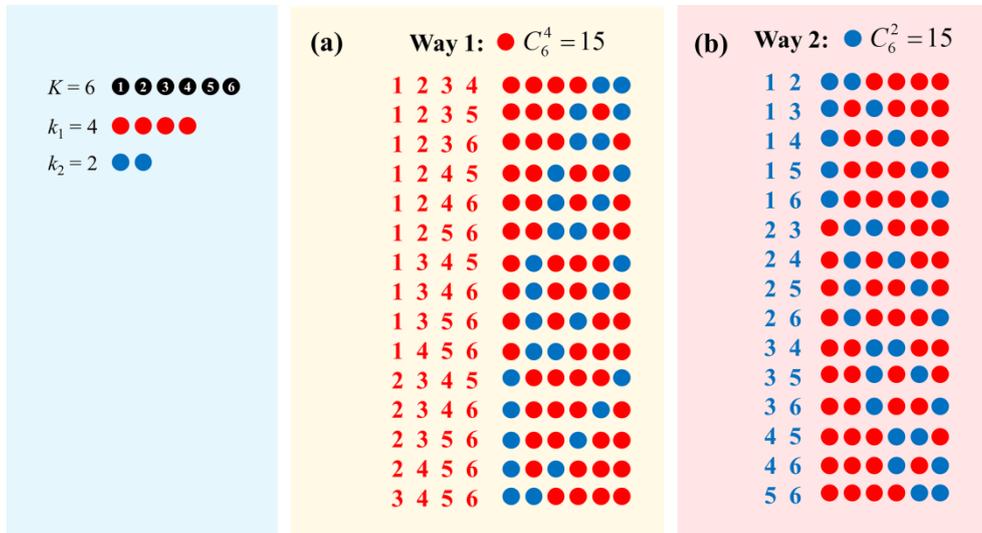

**FIG. 1**. (Color online) The two enumeration ways of binary site-occupancy configurations for $K = 6$, $k_1 = 4$ (red), $k_2 = 2$ (blue). (a) The first enumeration way, i.e. enumerating the configurations for red color, while the blue color occupying the remaining atomic positions. (b) The second enumeration way, i.e. enumerating the configurations for blue color, while the red color occupying the remaining atomic positions.

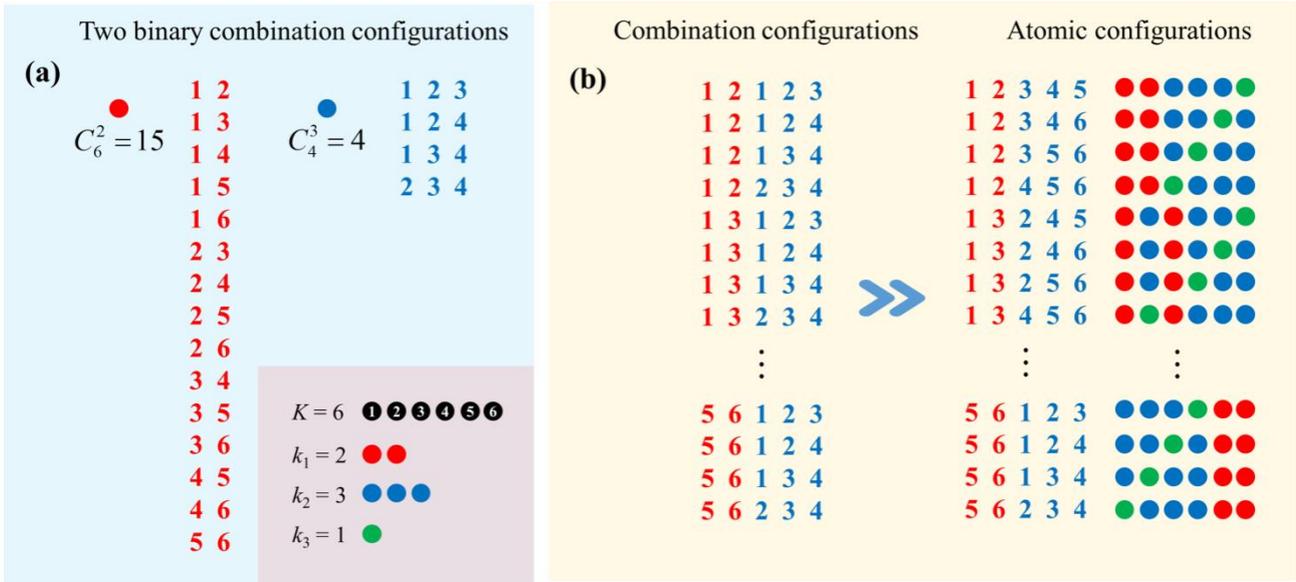

**FIG. 2**. (Color online) The enumeration of ternary site-occupancy configurations for $K = 6$, $k_1 = 2$ (red), $k_2 = 3$ (blue), $k_3 = 1$ (green). (a) The binary combination configurations for red and blue color. (b) The conversion of combination configurations to atomic configurations.

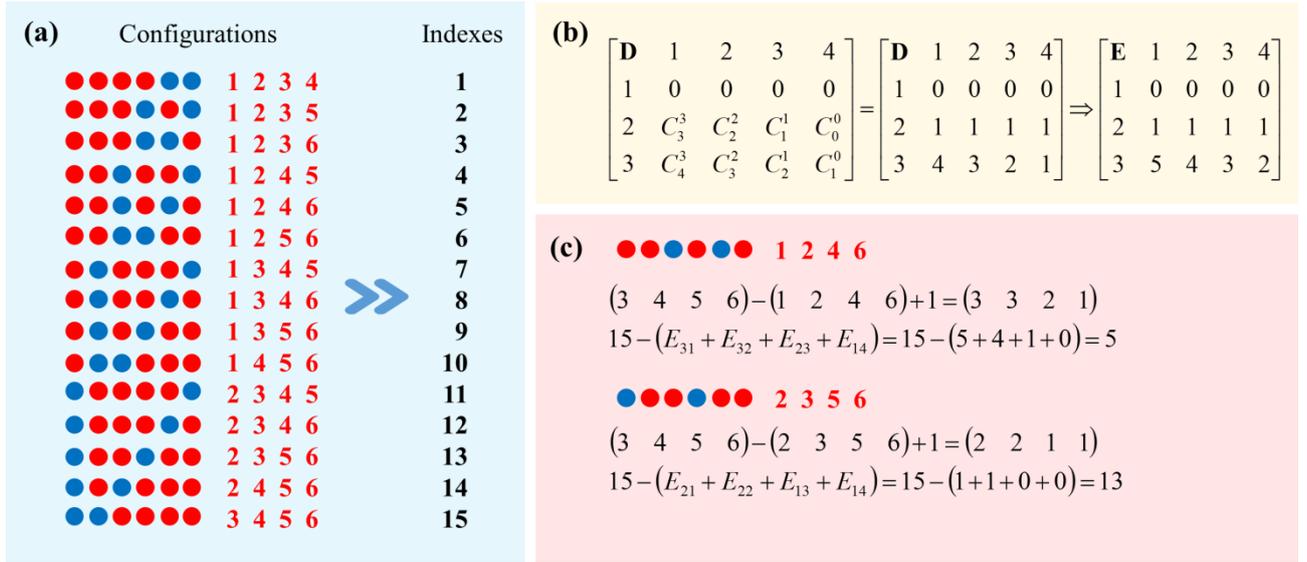

**FIG. 3**. (Color online) The conversion of binary site-occupancy configurations to consecutive integers (indexes) for $K = 6$, $k_1 = 4$ (red), $k_2 = 2$ (blue). (a) The atomic configurations and its corresponding indexes. (b) The constructing flow of matrix **E**. (c) Two specific examples of atomic configurations to integers.

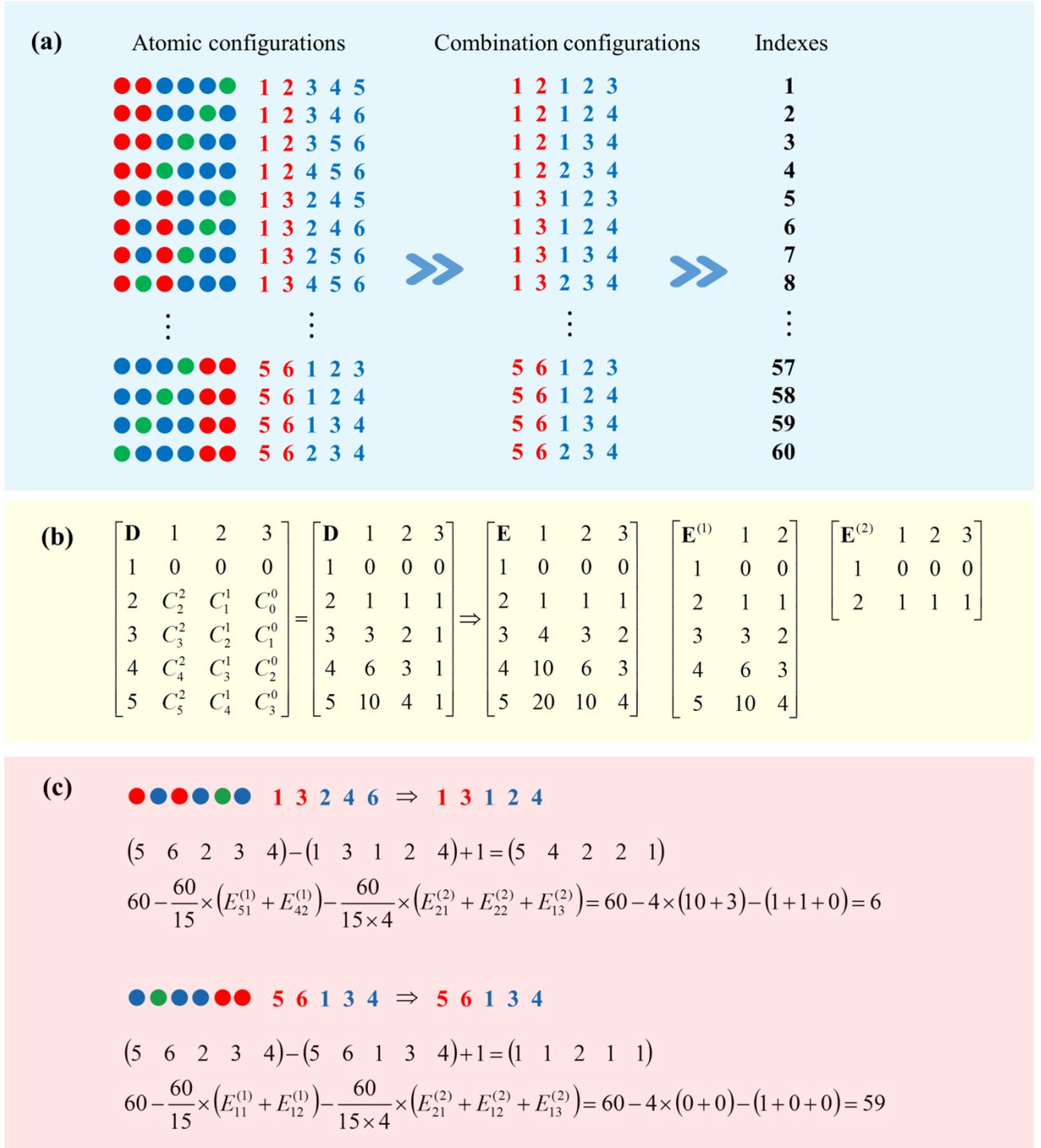

**FIG. 4**. (Color online) The conversion of ternary site-occupancy configurations to consecutive integers (indexes) for $K = 6$, $k_1 = 2$ (red), $k_2 = 3$ (blue), $k_3 = 1$ (green). (a) The atomic configurations and its corresponding combination configurations and indexes. (b) The constructing flow of matrix $\mathbf{E}$ and its sub-matrices $\mathbf{E}^{(1)}$, $\mathbf{E}^{(2)}$. (c) Two specific examples of atomic configurations to integers.

**FIG. 5**. (Color online) The algorithm flowchart of eliminating duplicate configurations.

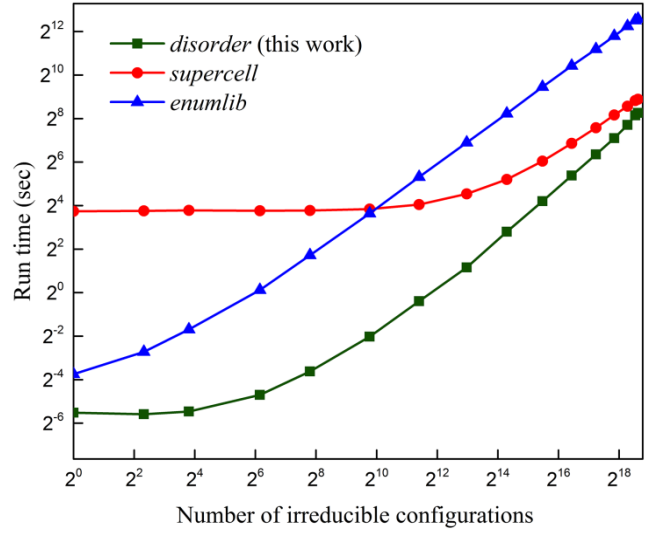

**FIG. 6**. (Color online) The run time of *disorder*, *supercell* and *enumlib* codes as a function of the number of irreducible configurations in a 2×2×2 FCC lattice. One can see that the *disorder* code (this work) has the lowest run time in the whole range of the number of irreducible configurations.